\documentclass[conference]{IEEEtran}
\IEEEoverridecommandlockouts
\usepackage{cite}
\usepackage{amsmath,amssymb,amsfonts}
\usepackage{algorithmic}
\usepackage{algorithm}
\usepackage[utf8]{inputenc}
\usepackage{graphicx}
\usepackage{textcomp}
\usepackage{dirtytalk}
\usepackage{booktabs, makecell}
\usepackage{etoolbox}
\usepackage{textcomp}
\usepackage{mathtools}
\usepackage{array}
\usepackage{xcolor}
\usepackage{multirow}
\usepackage{bbold}
\usepackage{siunitx}
\usepackage{comment}

\usepackage{subfigure}
\def\BibTeX{{\rm B\kern-.05em{\sc i\kern-.025em b}\kern-.08em
    T\kern-.1667em\lower.7ex\hbox{E}\kern-.125emX}}
\begin{document}

%\title{A Distributed Reinforcement Learning based Volt-Var Control for Inverter Based Resource in Power Distribution System \\

\title{Reinforcement Learning for Volt-Var Control: \\ A Novel Two-stage Progressive Training Strategy\\

\thanks{This project is supported by the DOE ARPA-E EDEGPRO project. 
%S.Zhang and N. Lu are with the Department of Electrical and Computer Engineering, North Carolina State University, Raleigh, NC, 27606 USA, (e-mail: szhang56@ncsu.edu, nlu2@ncsu.edu). )
}
}
\author{\IEEEauthorblockN{Si Zhang, Mingzhi Zhang, Rongxing Hu, David Lubkeman, Yunan Liu, and Ning Lu}
\IEEEauthorblockA{Department of Electrical and Computer Engineering} 
North Carolina State University, Raleigh, NC\\
\{szhang56, mzhang33, rhu5, dllubkem, yliu48, nlu2\}@ncsu.edu 
%\thanks{This  research  is  supported  by  the  U.S.  Department  of  Energy’s  Office  of Energy  Efficiency  and  Renewable  Energy  (EERE)  under  the  Solar  EnergyTechnologies Office Award Number DE-EE0008770.}
}

\maketitle

\begin{abstract}
This paper develops a \textit{reinforcement learning} (RL) approach to solve a cooperative, multi-agent Volt-Var Control (VVC) problem for {high solar penetration distribution systems}. The ingenuity of our RL method lies in a 
novel two-stage progressive training strategy that can effectively improve training speed and convergence of the machine learning algorithm.
%a scoring system for evaluating the performance of the control agents is developed using nodal Voltage magnitudes and reactive power injections as inputs. Next, the system-level rewards are computed by comparing the score between the controlled and non-controlled cases.} 
In Stage 1 (individual training), while holding all the other agents inactive, we separately train each agent to obtain its own optimal VVC actions in the action space: $\{\text{consume, generate,  do-nothing}\}$.
%Because the training of each agent can be carried out in parallel, the training time is significantly shortened. 
In Stage 2 (cooperative training), all agents are trained again coordinatively to share VVC responsibility. 
%to share the duty for voltage regulation. 
%As all agents are pre-trained in the first stage, the complexity for training multiple agents for coordination is  significantly reduced, which leads to fast convergence and robust performance. 
Rewards and costs in our RL scheme include (i) a system-level reward (for taking an action), (ii) an agent-level reward (for doing-nothing), and (iii) an agent-level action cost function. This new framework allows rewards to be dynamically allocated to each agent based on their contribution while accounting for the trade-off between control effectiveness and action cost.
The proposed methodology is tested and validated in a modified IEEE 123-bus system using  realistic PV and load profiles. Simulation results confirm that the proposed approach is robust and computationally efficient; and it achieves desirable volt-var control performance under a wide range of operation conditions. 

%the superior performance in terms of voltage regulating capability and progress of each stage. The proposed algorithm is robust and computationally efficient.

%a reinforcement learning based Volt-Var control algorithm for smart inverter in distribution system. Smart inverter(SI) en-able the PV to regulate the reactive power with wide range even when real power is limited by the radiance. This work first formulate the scoring system of Nodal Voltages and define the system-level voltage reward, which often used in centralized control fashion. Then a two-stage training strategy is proposed, where the each individual devices is trained individually in the first stage, and coordination among different devices are trained to allocate the contributions to each device in the second stage. 
\end{abstract}


\section{Introduction}

Solar photovoltaic (PV) systems equipped with smart inverters have superior continuous reactive power (Q) regulation capabilities compared with capacitor banks and voltage regulators. 
%Therefore, controlling PV inverters for providing reactive power is both fast and accurate. 
%The reactive power regulation can be decoupled with real power generation base on IEEE standard \textcolor{red}{what is the number of the standard? 1547} \cite{noauthor_ieee_nodate}.
Therefore, developing control strategies for distributed PV systems to provide Volt-Var control (VVC) is gaining increasing attention.  In general, there are three popular VVC approaches: rule-based, optimization-based, and more recently, machine learning-based. Although rule-based approaches are widely used in the field due to the ease of implementation, they lack the ability to adapt  to fast-changing operational conditions. The major drawbacks of optimization-based approaches are their strict requirement of accurate network models and complex computational platforms for implementation. Futuremore, the computational complexity increases exponentially as the system scale (e.g. number of controllable devices) increases. 

Machine learning, especially \textit{reinforcement learning} (RL), has been proven effective to generate optimal voltage control policies via offline and  online training [1-3].
%\textcolor{red}{give a ref. }. %The implementation of RL-based VVC controllers can be as easy as that of the rule-based ones with comparable control performance to the optimization-based VVC.  
Comparing to conventional rule-based VVC controls, main advantages of the RL-based VVC are its ease of implementation and high adaptability in a fast-changing operational environment.  Zhang \emph{et al.} \cite{zhang_deep_2021} and Sun and Qiu \cite{sun_two-stage_2021-1} proposed  multi-agent reinforcement learning (MARL) solutions for training VVC agents in both centralized and decentralized environments. 
%\textcolor{red}{??? What's the role of the next sentence? YL.} 
However, under this setting, the decentralized agent does not have learning capability - it only executes.
%\textcolor{red}{is this statement correct?}
%When communication is interrupted, the agent cannot adapt to new operation conditions via on-line learning.  -- I would discuss this in more detail in journal paper to save some space.} 
%\textcolor{orange}{This means the does not have individual learning capbability,decentralized execution does not have individual learning capbability, when there is communication loss, the centralized is down, and can not do on-line learning} 
Wang \emph{et al.}  \cite{wang_data-driven_2020} formulated the VVC problem as Markov game for solving the Voltage Violation problem using (one-shot) static environmental data as an episode. In order for agents to evolve their policies in response to a nonstationary environment, Lowe \emph{et al.}. \cite{lowe_multi-agent_2020}  developed a multi-agent deep deterministic policy gradient (MADDPG) method.
% that extended the above-mentioned framework to dynamic data series.  
However, inferring other agents' actions requires training additional neural networks, causing the design of VVC increasingly complex when the number of VVC agents increases.
%\textcolor{red}{??? Logical flow is strange here?? Yunan} \textcolor{green}{???si, I rewrote this a little but don't know if this is what you mean. Ning}\textcolor{yellow}{Yes, this is correct.}
%this difficulty can be mitigated by assuming the other actions of last time step is observable for all agents. \textcolor{red}{ what is the disadvantage of this paper you are solving in our proposed method?}

%\textcolor{red}{You literature review is insufficient.  1) We need to list a couple of RL papers (can be in other fields) to explain its advantage over other machine learning based or optimization based methods.  2) we need to reference to other RL papers in our filed to explain the state-of-the-art.  3) we need to point out what haven't be solved and what we can solve in our paper.  
%\textcolor{red}{I think the problems of the state-of-the-art is the increased training complexity when the number of VVC agents increase, the training speed is slow and convergence of RL is not guaranteed, and there lack of an effective scoring system and reward allocation system for training agents to collaborate with each other.  }

Centralized RL-based control design approaches often suffer from the so-called \textit{curse of dimension}. 
%MARL, a potential solution to this issue, 
Convergence and stability in training are usually difficult to achieve when many agents need to coordinate their operations in a fast changing environment. For example, an common scenario that often occurs in VVC training is passing clouds accompanied by rapid load changes in a distribution circuit with many PV systems.

To address the aforementioned issues, in this paper, we develop a two-stage RL approach to train multiple VVC agents progressively on a distribution feeder. Our contributions are two-fold. \emph{First}, we propose a novel reward design and allocation mechanism to account for the contributions of all agents; we aim to trade-off between control effectiveness and cost. In particular, aach VVC agent can take one of three basic actions: \say{generate-Q}, \say{consume-Q} and \say{do-nothing}.
%controlling nodal Voltage magnitudes within the operation limits 
The system's performance score is calculated by the degree of system-wide voltage violations for assessing VVC performance achieved by all VCC agents. Immediate reward is defined as the score of take-an-action (i.e., generate or consume ) minus the score of do-nothing. At the agent-level, the action cost is calculated according to the efforts committed by an agent.  %Thus, agent-level do-nothing rewards encourage agents to take no action and system-level awards encourage agents to take an action. 
The \say{do-nothing} reward allows us to include \say{do-nothing} as a \say{wise} action when the value of take-an-action diminishes. Note that the agent-level reward plays an important role in a decentralized, co-operative training environment. Rewards are not shared uniformly among all agents but rather dynamically assigned according to their efforts, which takes into account an agent's contribution while considering the cost for taking an action.
%This also properly reward the VVC agent for taking one of the three basic actions: "generate-Q", "consume-Q" and "do-nothing". 

\emph{Second}, we propose a novel two-stage, progressive training strategy. In Stage 1 (individual training), each agent is learn to take three basic control actions: \say{generate-Q}, \say{consume-Q}, and \say{do-nothing}, assuming all other agents are inert. Because the training of the agents can be conducted in parallel, the training time is unaffected by the number of agents. In absence of interventions from the other agents' actions, the agent currently being trained can focus on selecting one of the three actions with a fixed $Q$. This guarantees that our algorithm converges fast and is robust. In Stage 2 (cooperative training), 
%all pre-trained agents are trained together to learn "coordination", i.e., how much Q to deliver or absorb in the presence of other agents. 
as all agents have gained understanding of when to \say{generate-Q}, \say{consume-Q}, and \say{do-nothing}, the training can now focus on learning the ``optimal" magnitude of $Q$ an agent needs to provide in the presence of the other agents, i.e. learning coordination only. 
Thus, the training complexity is significantly reduced. Our results show that this 2-stage, progressive training approach is computationally much more efficient than the state-of-the-art methods, leading to faster convergence and more robust performance.

\vspace{-0.02in}

\section{Problem Formulation} \label{section2}
%In this section, we present model assumptions, reward structure, and formulations of our problem; we also describe the two-stage progressive training strategy for the RL-based multi-agent VVC.
\vspace{-0.03in}
\subsection{Assumptions} \label{assumptions}
\emph{First}, all actions are taken in fast control intervals (i.e., at 1- or 5- minute), so they are immediately observable to all VVC agents at time $t$. We can use the persistence model instead of policy inference to predict the other agents' actions by assuming that observations of the environment at $t-1$ are sufficient to predict the states of the environment at $t$.
%Second, we assume that when a robust policy is establishing, the policy changes slower than the environment. This is because a policy maps observations of an environment to actions.So when the control interval between two consecutive control and observation intervals is short, the actions suggested by the policy are similar majority of time. Therefore, policy inference for predicting other agents' actions is no longer needed. 
\emph{Second}, the communication among agents is via the system operator, who is responsible for letting a VVC agent ``know" of required information at time $t$ (e.g., actions taken by all other agents at $t-1$). However, the other agents' actions at $t$ are hidden (and only revealed at the next step) so each agent will make its own decision independently. The parameters of the policy network of a VVC agent are also unknown to other agents. 
\emph{Third}, there is no other VVC devices on the feeder so that the PVs are the only resources for reactive regulation. \emph{Fourth}, the only objective of an VVC agent is to control the nodal voltages to be within the defined operational range.  

\vspace{-0.05in}
%\subsection{State Space and Action Space}
\subsection{Problem Formulation}
To formulate the VVC problem as a Markov decision process (MDP), in that we define the global state, $\mathcal{S}^{t}$, partial observation $\mathcal{O}_i^{t}$, and action set, $\mathcal{A}$, of the $i^{\mathrm{th}}$ VVC agent at time $t$ as:
\begin{eqnarray}
\mathcal{S}^{t} &:=&
\begin{bmatrix}
\mathcal{V}^{t} & \mathcal{P}_{\mathrm{pv}}^{t}& P_{\mathrm{feeder}}^t& Q_{\mathrm{feeder}}^t
\end{bmatrix} \label{eq1a}\\
\mathcal{O}_{i}^{t} &:=&
\begin{bmatrix}
\mathcal{V}^{t} & P_{\mathrm{pv},i}^t& P_{\mathrm{feeder}}^t& Q_{\mathrm{feeder}}^t& \mathcal{A}^{t-1}
\end{bmatrix} \label{eq1b}\\
a_i^t&:=&Q_{\mathrm{pv},i}^t  \label{eq1c}\\
\mathcal{A}^{t-1} &=&
\begin{bmatrix}
{a}^{t-1}_1\dots {a}^{t-1}_i\dots a^{t-1}_N
\end{bmatrix}, \quad   \forall i \in [1,N] \label{eq1d}\\
\mathcal{A}^{t} &=&
\begin{bmatrix}
{a}^{t}_1\dots {a}^{t}_i\dots a^{t}_N
\end{bmatrix}, \quad  \forall i \in  [1,N] \label{eq1e}\\
\mathcal{V}^{t} &=&
\begin{bmatrix}
{V}_{1}^t\dots {V}_{k}^t \dots V_{M}^t
\end{bmatrix}, \quad  \forall k \in [1,M]\label{eq1f}
\end{eqnarray}
%At time step $t$, the state $\mathcal{S}$(input information) of agent $i$ can be expressed as \eqref{eq3}.
where $\mathcal{V}^{t}$ is the nodal voltage set; $\mathcal{P}_{\mathrm{pv}}^{t}$
%in \eqref{eq_add} 
is the active power output Set of PV farms at step $t$; $P_{\mathrm{pv},i}^t$ is the $i^{\mathrm{th}}$ PV real power output at $t$, $P_{\mathrm{feeder}}^t$ and $Q_{\mathrm{feeder}}^t$ are the active and reactive power output at the feeder head $t$, respectively; 
${a}^{t}_i$ is the action taken by the $i^{\mathrm{th}}$ agent for generating (positive) or consuming (negative) reactive power of $Q_{\mathrm{pv},i}^t$; $M$ is the number of nodes being monitored; $N$ is the number of VVC agents. Note that $\mathcal{S}^{t}$ represents the global view of the environment and $\mathcal{O}_{i}^{t}$ describes the agent's local view of the environment. %--- I rewrite your sentences based on my reading of the formulation - Is this statement correct?}\textcolor{blue}{Yes, it is correct.}

Note that $\mathcal{A}^{t-1}=0$ when $t=1$. The action space of a centralized VVC controller is an action set because all agents' actions need to be considered. However, the action space of the $i^{\mathrm{th}}$ distributed VVC agent is a scalar, which is the reactive power output of the $i^{\mathrm{th}}$ PV farm, $Q_{\mathrm{pv},i}^t$ (see \eqref{eq1c}).

\subsection{VVC Performance Score Calculation}
\begin{figure}[htbp]
\vspace{-0.1in}
\centerline{\includegraphics[width=.495\textwidth]{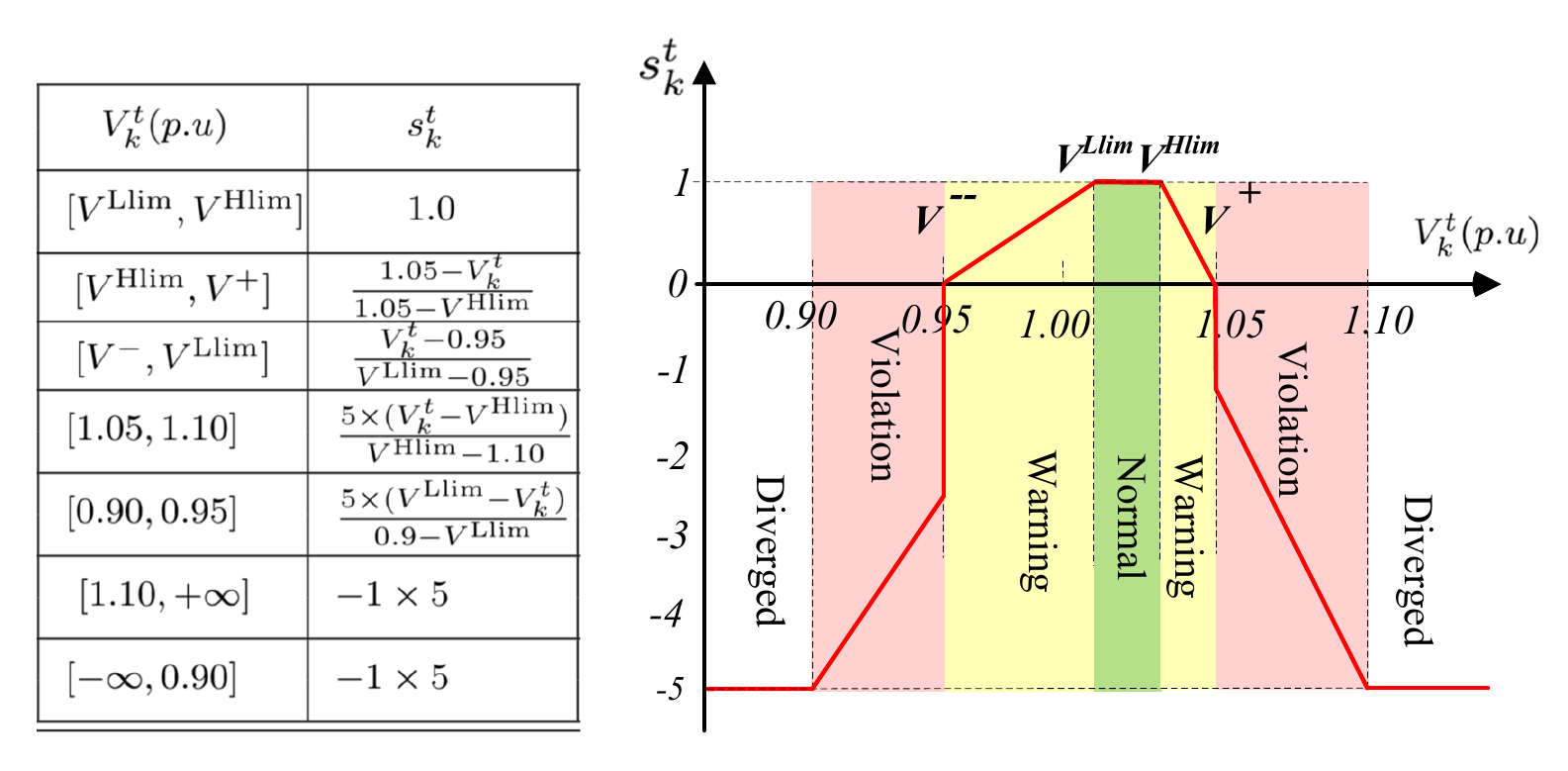}}
\vspace{-0.1in}
\caption{The setup of the nodal voltage score curve.}
\label{fig1}
\vspace{-0.05in}
\end{figure}

The ANSI standard requires the distribution system voltage to be maintained within the interval $[V^{\mathrm{-}}, V^{\mathrm{+}}]$, with $V^{\mathrm{-}}=0.95$ and $V^{\mathrm{+}}=1.05$ p.u. However, a utility may choose to hold system voltage to be within another designated interval  $[V^{\mathrm{Hlim}}, V^{\mathrm{Llim}}]$. Inspired by \cite{wang_data-driven_2020}, we revised the control target from a single voltage reference to a set of piece-wise linear score functions, as shown in Fig. \ref{fig1}. %Three pairs of thresholds are selected to divide the score curve into seven voltage zones, where 
Note that $s_k^t$ is the voltage score calculated for node $k$ at time $t$. 

%By scoring all the nodes inside the feeder, we can globally evaluate the circuit Voltages. 
%Because an agent does not have prior knowledge of which nodal voltage violations it should regulate, the system-level score is calculated as \textcolor{red}{I don't understand what you are doing here.  I think the system level score should be the sum of all nodal scores. You can use an average, but this has nothing to do with agents.}
%the average score of all the nodes. Due to the average factor in \eqref{eq5}, the penalty scale set as 5.

The system score, $Score_{A}^{AS}$, is defined as the average nodal voltage score, that is, 
 \begin{equation}
    Score_{A}^{AS}= \frac{1}{M}\sum_{k=1}^{M}{s^t_{k}},\label{eq5}
\end{equation}
where $AS$ is the joint action space and $A$ is the joint action set. Note that the score curve is capped at 5.0 after the system voltage drops below 0.9 p.u. or surpasses 1.10 p.u.

\subsection{Design and Allocation of Reward}

The system-level reward $r_{s}$ is defined as 
\begin{equation}\begin{split}
    r_{s} := Score_{A}^{AS}- Score_{DN}^{AS},\label{eq6}\end{split}
\end{equation}
where $DN$ indicates the action of \say{do-nothing}. 

Our reward definition follows the idea of the \textit{Advantage Actor Critic} (A2C) method \cite{mnih_asynchronous_2016}.  We deduct the actual reward from a baseline to reduce the variance of policy gradient so that policy network can be trained easily. The baseline score can be computed from (\ref{eq5}). This advantage reward can be explicitly formulated and meaningful to show the effectiveness of taking actions.

In the single-agent setting, the agent can seek effective actions by bench-marking against a predetermined baseline action. %\textcolor{red}{benchmarking against a predetermined baseline action.} %\textcolor{red}{YL:what's the meaning of `unified'? NL: yeah I had the same question. Do you mean "the same baseline action"}\textcolor{orange}{Yes, I mean the same baseline actions along the time, for example "Generate Q".}. 
However, when there are many VVC agents in the system, the dimensionality of the (joint) action space increases drastically. In addition, intervention between actions taken by different agents makes the baseline action selection much more complicated. For simplicity, we use the action of \say{do-nothing} as a unified baseline action in multi-agent setting. In power distribution systems,  \say{do-nothing}, in absence of voltage violations, is indeed often preferred as a baseline action for controlling a VVC device.

\section{two-stage Progressive Training}
%As mentioned the baseline actions of reward definition, we can not directly have multiple baselines as comparison in multi-agent setting other than "Do Nothing". But for single agent setting, we could easily get such a baseline action better than "Do Nothing". The proposed 2-stage training approach leverage this features to improve the sample efficiency. In the first stage, each agent is individually trained to acquire policy to make basic operation decision. In the second stage, every agent will based on basic operation to learn to coordinate with other agents.
\subsection{Stage 1: Individual Training}
The goal of the first stage is to train a VVC agent by considering two simplified basic control strategies: i) when the voltage violation cannot be alleviated by it action, do-nothing, and ii) when taking an action, what is the polarity of the action, i.e., generate (+) or consume (-). 

The agent-level reward for the $i^{\mathrm{th}}$ agent in stage 1, $r_{1,i}$, when taking an action, $a^t_i$, is expressed as
\begin{eqnarray}
    Cost_i &=& w_{\mathrm{cost}} \times |Q^t_{i}| \label{eqn4}\\
        r_{DN}&=&
        \left\{ \begin{array}{ll}
           \num{ 1e-3} &|a^t_i|\leq a^{\mathrm{th}} \\
            0 &\text{otherwise}
        \end{array} \right. \label{eqn5}\\
r_{1,i} &=& r_{s}- Cost_i + r_{\mathrm{DN}}\label{eqn6}
\end{eqnarray}
where $Cost_i$ is the \say{cost} for taking an action for the $i^\mathrm{th}$ agent, $w_{\mathrm{cost}}$ is the weighting factor for computing $Cost$  from $|Q^t_{i}|$, $a^{\mathrm{th}}$ is the action threshold, and $r_{\mathrm{DN}}$ is the do-nothing reward. 

% \textcolor{cyan}{ This alignment is ok but the spacing between variables looks wired. You can also try to align the starting points of the equation.
% A way to align the equations:
% \begin{eqnarray}
%         Cost_i &=& w_{\mathrm{cost}} \times |Q^t_{i}| \label{eqn4}\\
%         r_{DN} &=& \left\{ \begin{array}{ll}
%           \num{ 1e-3} &|a^t_i|\leq a^{\mathrm{th}} \\
%             0 &\text{otherwise}
%         \end{array} \right. \label{eqn5}\\
%         r_{1,i} &=& r_{s}- Cost_i + r_{\mathrm{DN}}\label{eqn6}
% \end{eqnarray}
% }

Note that \eqref{eqn6} aims to trade-off between reward and cost. The $r_{\mathrm{DN}}$ term can not only encourage the agent to take no action while the effectiveness of its effort diminishes, it also forms a non-action zone %\textcolor{red}{what's ``dead-band"? rephrased. 
to avoid unnecessary oscillatory actions. The above setting of the reward structure constitutes one of the major novelties of our approach.% We consider this is a major contribution in the RL reward mechanism design.

%\textcolor{red}{no need to put $r^t_i$ if you do not keep a time record there for the i VVC agent.}

When regulating nodal voltages on a distribution feeder, each VVC agent has an effective control range determined by the network topology and the location of the PV farm. Thus, when an agent's action is ineffective or only marginally effective for mitigating voltage violations, the optimal strategy is do-nothing. However, when multiple agents are being trained jointly in a nonstationary environment, two main challenges arise: i) the lack of appropriate baseline actions for assessing performance improvements and ii) the lack of a fair performance-driven reward allocation mechanism for each agent.  Consequently, the training process becomes lengthy and unstable. Convergence to the optimal VVC control and coordination strategy for all agents is therefore difficult to achieve. 

\textbf{Main advantages.} If there exists only one VVC, the training converges quickly. This is because the radial distribution network topology ensures a relatively linear V-Q relationship. Furthermore, as the agent receives full credits/penalties for its action as specified in (\ref{eqn6}), learning the polarity of an action is straightforward. As the first stage training can be conducted in parallel, having multiple agents will not slow down the training process. After the first-stage training, all agents should ``understand" when their actions are effective. This is a very important feature for reward allocation in the second-stage training, because all action-taking agents are considered effective contributors. Through an appropriate credit-sharing mechanism, agents can learn to contribute the right amount of $Q$ in presence of other agents.

%extract the inherent impact of the circuit topology to scores increase, then the agents will decide the basic operation (polarity): Generate, DoNothing or Consume. In this stage, the key consideration is to pick the feasible basic operation which could lead to significant score increase. Therefore, do-nothing reward as incentive, as well as cost term, are formulated in this stage. The last time actions are set zeros for \eqref{eq3} in this stage.

% \textit{Cost Term}: From system reward definition, it mainly covers the effectiveness of taking actions. Given that controllable resources are sometimes redundant, and each agent have flexibility than system needs, we intend to avoid “unnecessary” actions. The cost term of each agent is introduced for agent to evaluate the “effectiveness” of their actions in \eqref{eq8}, where $C_{w}$ is the cost weight, and since the var regulated capability is directly related to its capacity so all $Q$ in this paper is the refer to "per unit" value.
% \textit{Do Nothing Reward}: The Do-Nothing Reward is fixed reward when the agents choose the basic operation of “Do Nothing”. This term is stimulate the agents to do nothing by default if they do not gain much reward by taking other actions, described in (\ref{eq_add3}), where $r_{DN}$ is the Do Nothing reward and $a_{th}$ is the action threshold.
%In this stage, since each agent is training separately, the system level reward attributes to the agent actions. The relationship between agent action reward $r(a_i)$and system level reward $r_s$ in stage 1 is formulated as \eqref{eq11}.

\subsection{Stage 2: Cooperative Training}
The goal of the second stage is to train all VVC agents jointly in the same environment so that each VVC agent can learn to \emph{ generate/consume its own share} of reactive power when coordinating with the other agents for reducing nodal voltage violations. 
%Taking other agents' actions into consideration where the basic operation is provided in the first stage. 
Our assumption is that, after stage 1, each agent has learned to take only effective actions, i.e., an agent will be inert when its action will not help to alleviate voltage violations and will know when to generate/consume $Q$. As defined in (\ref{eq1b}), in the second stage, the actions taken by all agents, $\mathcal{A}^{t-1} $, at $t-1$ will serve as an input for the observation space of all VVC agents at $t$. 

In the second stage, the agent-level action reward, $r_{2,i}$, is calculated as
\begin{equation}
\begin{split}
 r_{2,i}& := CF_i\times r_{s} - Cost_i   \label{eq7}
\end{split}
\end{equation}
where the cost is calculated using (\ref{eqn4}) and the contribution factor $CF_i$ can be calculated as
\begin{equation}
     CF_i := \frac{|Q^t_{i}|}{\sum_{j=1}^{N} {|Q^t_{j}|}}
     \label{eqn8}
\end{equation}
Note that here we use the absolute value of $Q$ because there may be cases in which one agent is generating $Q$ for boosting its local voltage while another agent is consuming $Q$ for suppressing its local voltage. In this case the two agents are collaborating with each other to remove the voltage violations. 
 %where $a_{th}$ is threshold for environment, 
Let $a_{\mathrm{S_1},i}$ and $a_{\mathrm{S_2},i}$ be the action outputs by the stage-1 policy network, $\mathrm{S_1}$, and stage-2 policy network, $\mathrm{S_2}$ by agent \textit{i}, respectively. The agent's final action, $a_{2,i}$, is determined by 
\begin{equation}
\begin{split}
 a_{2,i} & := sign(a_{\mathrm{S_1},i})\times\mathbb{1}(|a_{\mathrm{S_1},i}|>a_{th})\times a_{\mathrm{S_2},i}  \label{eqn9}
\end{split}
\end{equation}
Note that (\ref{eqn9}) generates desired control strategies at two levels: i) Stage-1 policy network determines a ``raw" action: whether an action is needed and if so, its polarity; ii) Stage-2 policy network provides a ``complete" action by prescribing the magnitude of the action when such an action is needed.

%In this stage, the focus is coordination training for all agents. 
%A challenge in the multi-agent setting is, insufficient coordination among agents can lead to over- or under- compensation of voltage violations. 

 %\textit{Reward Allocation}:
 %A reward allocation calculation are formulated to provide a transformation from system-level reward to agent-level rewards by providing each agent with $CF (Contribution Factors)$. It represents the participation factor of the $a_i$ for the whole system among all other actions. It is the key concept to simplify from Markov Game joint reward $r(\mathcal{S}^{t},\mathcal{A}^{t})$(or system level reward $r_{s}$) to distributed $r(\mathcal{O}_{i}^{t},a_i^t)$. Note that the sign of the basic operation are determined in the first stage, the $CF$s calculated in this stage are all non-negative values.

%With the reward allocation formulation, 
\subsection{Algorithm Implementation}
\begin{figure}[htbp]
\centerline{\includegraphics[width=.45\textwidth]{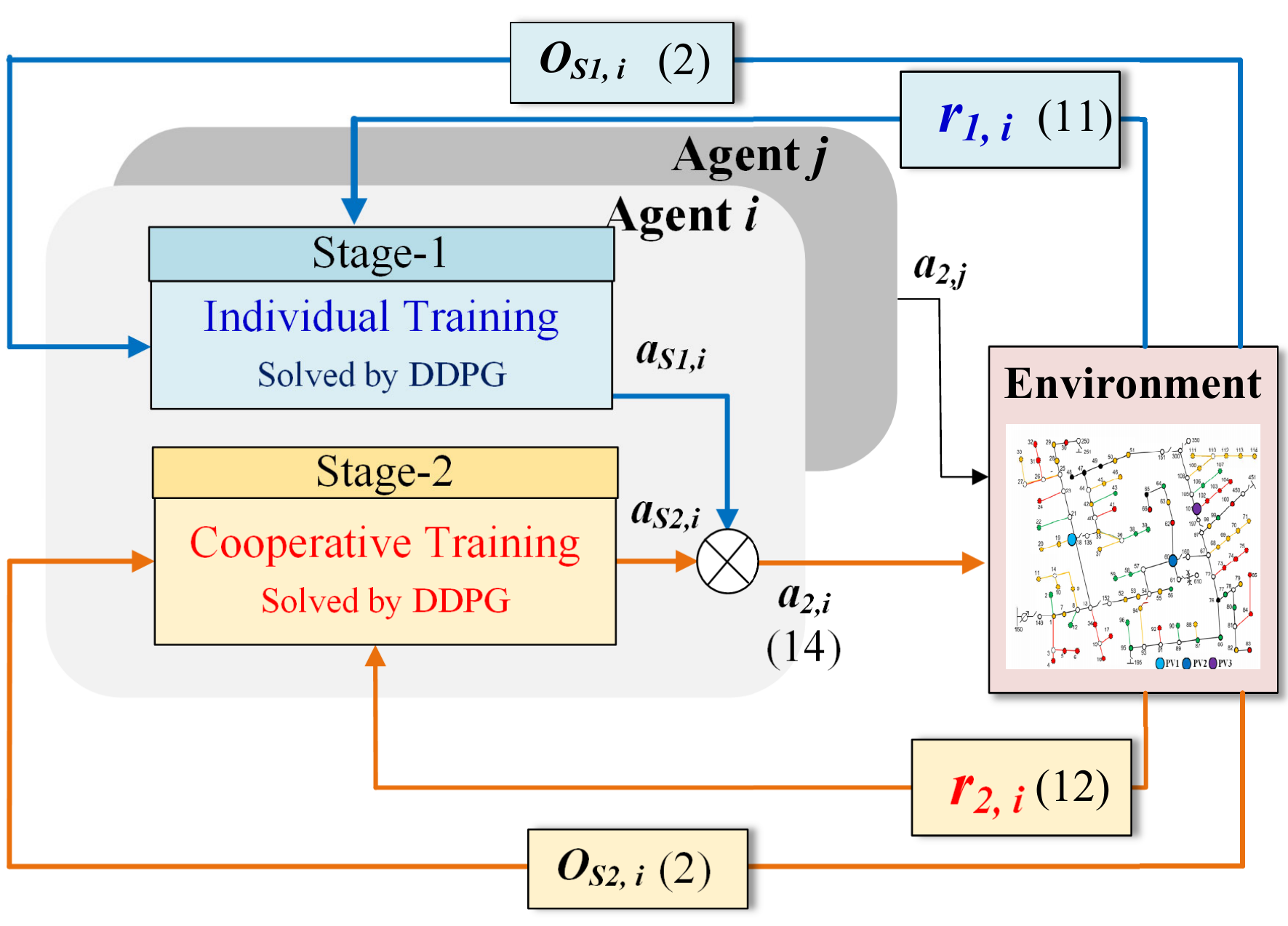}}
\vspace{-0.05in}
\caption{An illustration of the two-stage training process using DDPG.}
\label{fig4}
\vspace{-0.1in}
\end{figure}

We solve the  distributed RL problem using DDPG proposed in \cite{lillicrap_continuous_2019} following the workflow shown in Fig. \ref{fig4}. Note that $\mathcal{O}_{\mathrm{S1},i}$ and $\mathcal{O}_{\mathrm{S2},i}$ are the partial observations to agent $i$ in stages 1 and 2, respectively. In (\ref{eq1b}), for $\mathcal{O}_{\mathrm{S1},i}$, $\mathcal{A}^{t-1}=[a^{t-1}_i]$  since there is no other agent in the system; for $\mathcal{O}_{\mathrm{S2},i}$,  $\mathcal{A}^{t-1}=[a^{t-1}_1... a^{t-1}_i...a^{t-1}_N]$, given the second assumption in Section \ref{assumptions}.
\section{Numerical Studies}
\begin{figure}[htbp]
\centerline{\includegraphics[width=.45\textwidth]{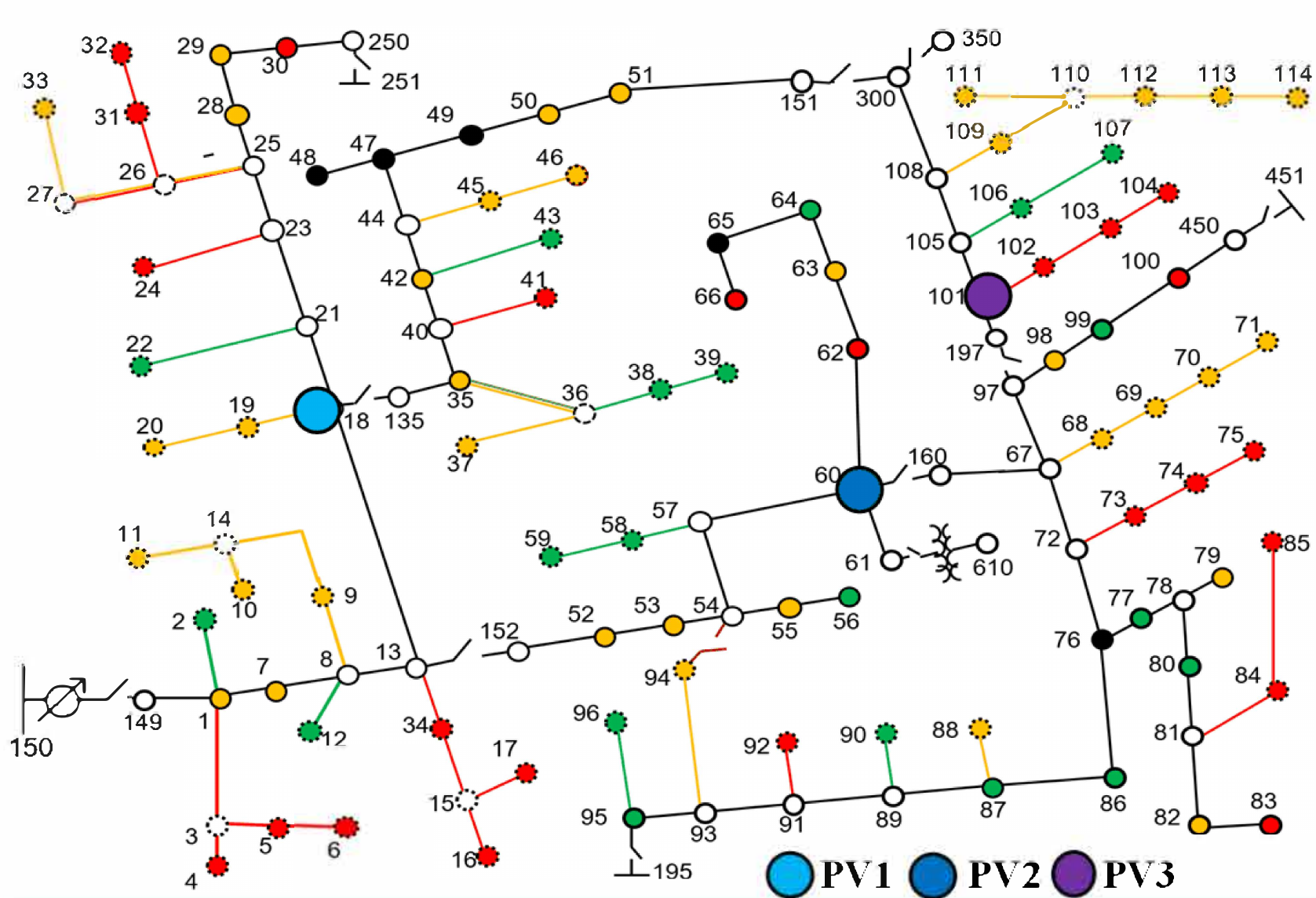}}
\vspace{-0.1in}
\caption{Configuration of the training environment. (Test feeder topology: modified IEEE 123-bus syste. Green, red, blue, and black lines: $a$, $b$, $c$, and 3-phase circuits, respectively. Empty circles: buses without loads.)}
\label{fig2}
\vspace{-0.1in}
\end{figure}

The training is conducted on a testbed developed using the topology of the IEEE 123-bus system, as shown in Fig. \ref{fig2}. The back-end of the environment is OpenDSSDirect running on Python. The RL agents are trained using Pytorch. The annual load and PV data are generated from the PECAN street data set \cite{pecan_street}. We consider a 5-minute control interval and a 30-minute learning episode.  Table \ref{tab2} lists the locations and capacitys of all PV farms.  All PV inverters are oversized so $S_{pv} = 1.08 P_{pv}$. According to \cite{noauthor_ieee_nodate}, the inverter regulates $Q$ within $[-44 \%, 44 \%]$ of the PV rated capacity, $S_{\mathrm{pv}}$. Voltage regulators set at 1:1 ratio mode and is inert during the training. The utility preferred voltage operation range is determined by $V^{\mathrm{Hlim}}=1.03$ p.u. and $V^{\mathrm{Llim}}=1.01$ p.u., the values of which are within the ANSI limit.
\begin{table}[htbp]
\scriptsize
\caption{Location and capacity of the PV farms}
\vspace{-0.15in}
\begin{center}
\begin{tabular}{cccc}
\hline
\hline
\textbf{Inverter}&\textbf{Connected Bus(Phase)}&\textbf{Installment Capacity}\\
\hline
PV1 & 18(a,b,c)&800 kW\\
PV2 & 60(a,b,c) & 600 kW\\
PV3 & 101(a,b,c) & 300 kW\\
Total PV capacity &&1700 kW\\
\hline
\hline
\end{tabular}
\label{tab2}
\end{center}
\vspace{-0.15in}
\end{table}
\subsection{RL-based VVC Performance in four Seasons}
\begin{figure}[htbp]
\centerline{\includegraphics[width=.40\textwidth]{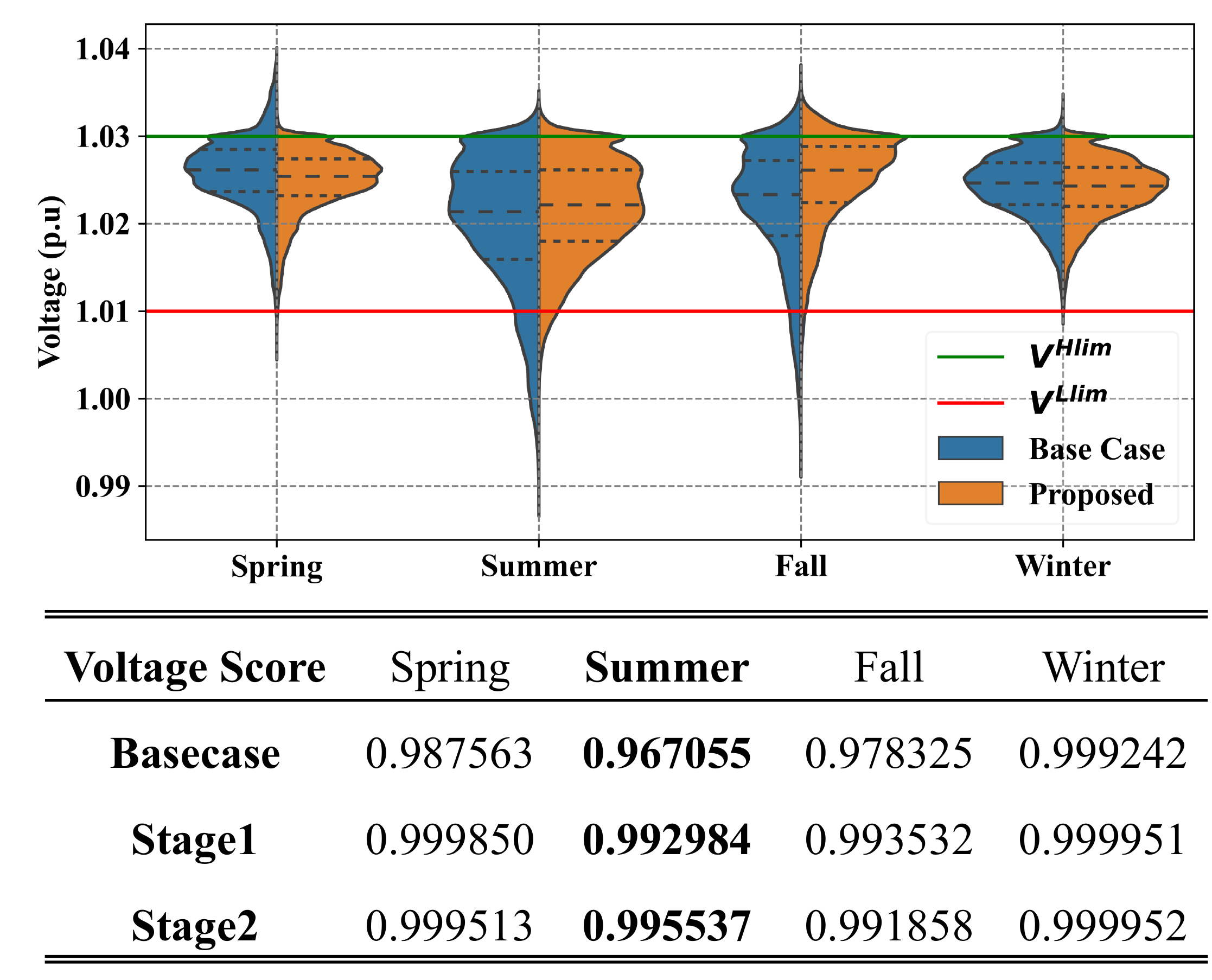}}
\vspace{-0.15in}
\caption{Distributions of nodal voltages and summary of voltage scores. }
\label{fig3}
\vspace{-0.10in}
\end{figure}

In each season, 20 days are selected for training and 3 days for testing. The base case is the \say{do-nothing} case. The nodal voltage distributions and the three-day-average voltage scores of the base case and the proposed VVC cases for the four seasons are summarized in Fig. \ref{fig3}. The left side of the violin plot represents the base case and the right side is the VVC case. The base case results show that in summer, the nodal voltages often drop below $V^{\mathrm{Llim}}$ (i.e. 1.01 p.u.) and in spring, the second worst, the nodal voltages often go above $V^{\mathrm{Hlim}}$ (i.e. 1.03 p.u.). As shown by the inner quartiles of the plots, the nodal voltage for the VVC case are significantly improved and most of the time the nodal voltages are within the preferred operation zone. As shown in Fig. {\ref{fig3}}, in winter, the voltage violations are rare so for the remaining studies, we only show the results obtained in the summer and spring. 
% \begin{comment}
% \begin{table}[htbp]
% \caption{System-level Voltage Score (3-day Average)}
% \centering
% \begin{tabular}{cccc}
% \hline \hline
% \textbf{Voltage Score}&\textbf{Basecase}&\textbf{Stage 1}&\textbf{stage 2}\\
% \hline
% Spring&	0.987563	&0.999850&	0.999513\\
% \textbf{Summer}&	\textbf{0.967055}	& \textbf{0.992984}&	\textbf{0.995537}\\
% Fall&	0.978325	&0.9935319 &	0.991858\\
% Winter&	0.999242	&0.999951&	0.999952\\
% \hline \hline
% \end{tabular}
% \label{tab3}
% \end{table}
% \end{comment}
\vspace{-0.15in}

\subsection{Performance Comparison with the Decentralized VVC}
\begin{figure}[htbp]
\centerline{\includegraphics[width=.36\textwidth]{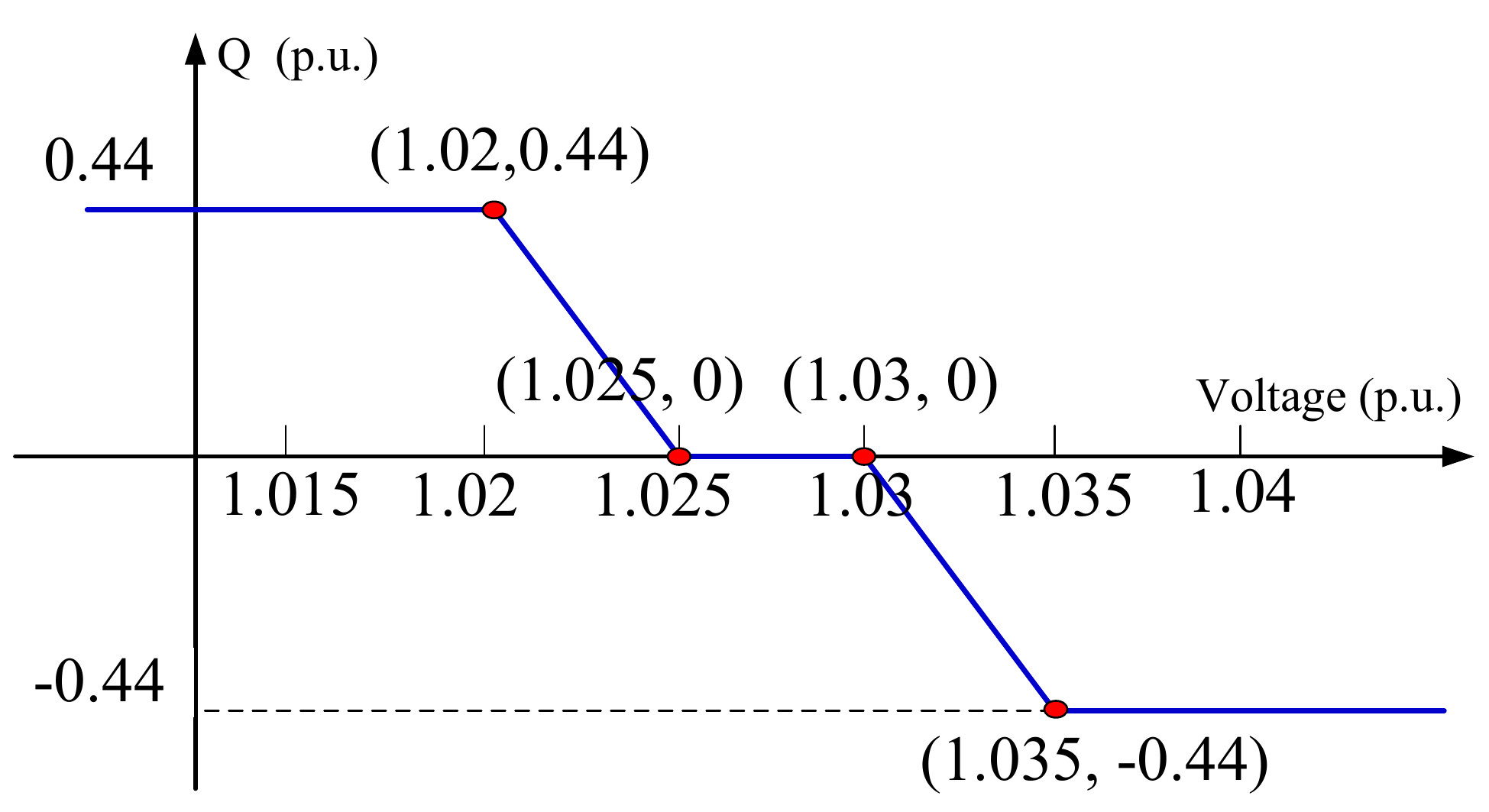}}
\caption{Conventional decentralized Volt-Var control curve.  }\label{figX}
\end{figure}

The parameters of a set of conventional inverter-based decentralized VVC control curves  \cite{noauthor_ieee_nodate} are shown in Fig. {\ref{figX}}. 
%\textcolor{red}{ --- you may need to draw the curve as it is hard to perceive what this look like. send you comment to wechat. Show those parameters in the figure. (1.02,0.44), (1.025,0), (1.03,0), (1.035,-0.44).} 
%can be set up . We carry out the performance comparison summer case. Note that the required band is tight, 
%The curve we choose is aggressive set. 
%The 3 pairs of separation points on the voltage score curve (See Fig. \ref{fig1}) ares 
As shown in Table \ref{tab4}, the conventional decentralized VVC takes the least number of actions, which is measured by the cumulative $Q$ consumption, $\sum Q$. However, it receives the lowest Voltage score, showing an inferior voltage regulation performance. Stage-1 policy does not consider coordination. Thus, PV1 always generates $Q$, causing more $V^{\mathrm{Hlim}}$ violations. Stage-2 policy has the highest voltage score, showing superior VVC control performance. By coordinating with other agents, $\sum Q$ is significantly reduced in stage-2.
%The voltage score and Q consume for different PVs are shown in Table \ref{tab4}, and Voltage boundaries are shown in the Fig.\ref{fig7}. The voltage and total reactive power distribution are presented in Fig.\ref{fig5}. We can see that t 

\vspace{-0.2in}
\begin{figure}[htbp]
\centerline{\includegraphics[width=.495\textwidth]{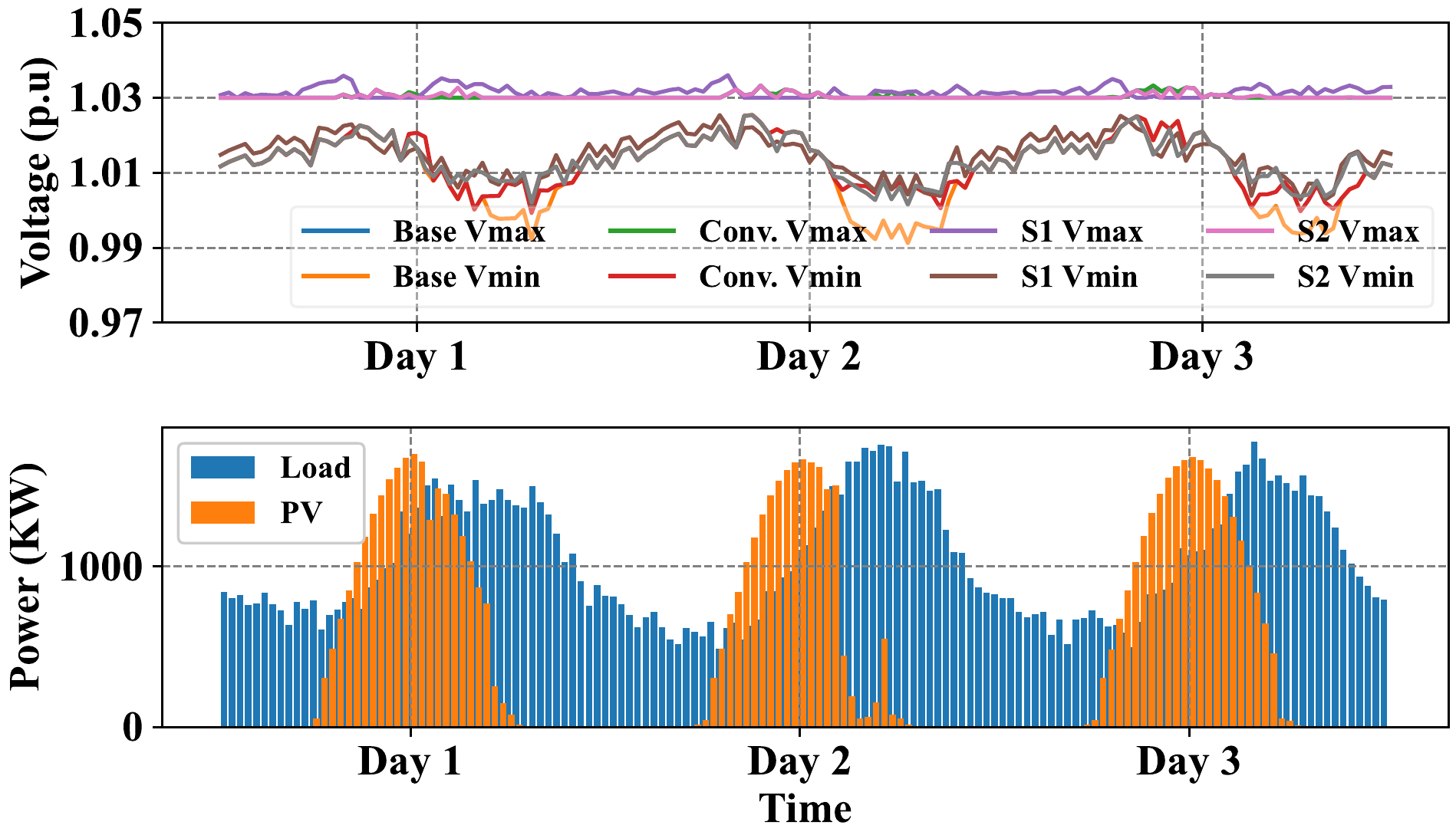}}
\vspace{-0.15in}
\caption{Voltage, PV, and load profiles in the three testing days. }
% \vspace{-0.10in}
\label{fig7}
\end{figure}
\begin{figure}[htbp]
\vspace{-0.10in}
\centerline{\includegraphics[width=.45\textwidth]{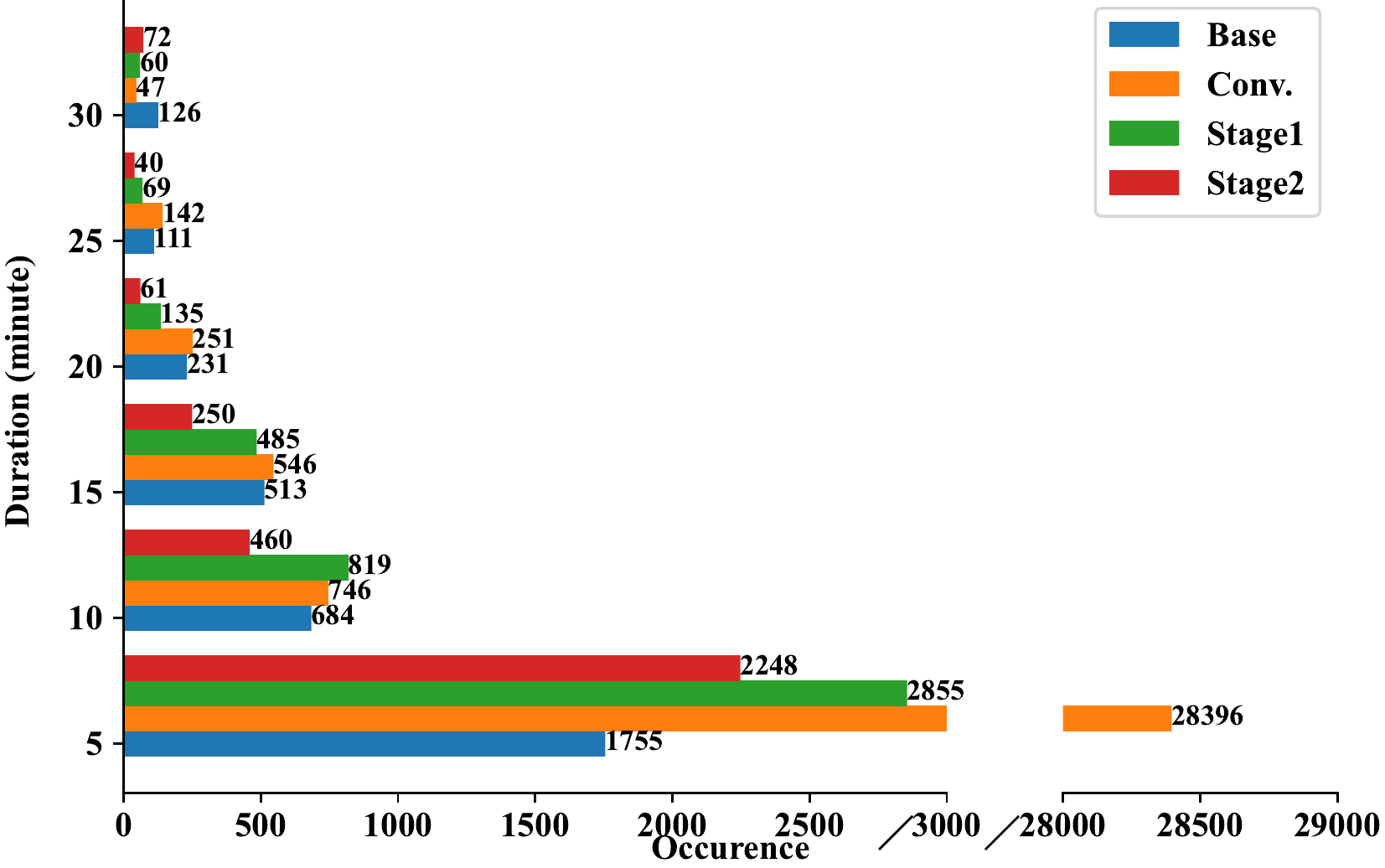}}
\vspace{-0.10in}
\caption{ Distribution of voltage event duration}
\label{fig9}
\vspace{-0.10in}
\end{figure}

% \begin{comment}
% \begin{figure}[htbp]
% \centerline{\includegraphics[width=9.2cm, height=4.8cm]{Voltage_Analysis.pdf}}
% \caption{Nodal Voltage Time Series Heatmap of Day2 of Stage2\textcolor{red}{why we need to show this figure?}\textcolor{blue}{I was think show nodal voltage information, since fig.6 just show Vmax and Vmin, but along with error information, I think this info is redundant, so may comment out.}}
% \label{fig8}
% \end{figure}
% \end{comment}
\begin{table}[htbp]
\scriptsize
\caption{VVC Performance Comparison (the summer Case) Evaluation}
\vspace{-0.15in}
\begin{center}
\begin{tabular}{cccccccc}
\hline
\hline
\multirow{2}{4em}{Algorithm}&Voltage&$Q_{total}$&$Q_{pv1}$&$Q_{pv2}$&$Q_{pv3}$\\
&Score& (kVAR)&$(p.u)$&$(p.u)$&$(p.u)$\\
\hline
Base Case&	0.98756	& - & -& -& -\\
Conventional&0.98995&93.859&0.01688 &0.07611 &0.09242  \\
Stage1&0.99286&452.43& 0.41250 & 0.12527&0.04583 \\
\textbf{Stage2}&\textbf{0.99556}&\textbf{144.03}&\textbf{0.01660} & \textbf{0.14361}& \textbf{0.11305}\\
\hline
\hline
\end{tabular}
\label{tab4}
\end{center}
\vspace{-0.15in}
\end{table}
 If some nodal voltages fall outside of the designated interval [$V^{\mathrm{Llim}}$$V^{\mathrm{Hlim}}$] in a control interval, we consider this interval to be a voltage violation event. Then, we compare the duration of such voltage events in four use cases: base case, conventional, stage-1 policy, and stage-2 policy in the summer season. Table {\ref{tab5}} and Fig. \ref{fig9} summarize the statistics of the durations of all voltage violation events  in the three summer testing days. 
%shows the occurrence of Nodal voltage warning of the summer case, 
%In the plots, the number of occurrence for voltage violation duration from 5 to 30 minutes. 
Conventional VVC is effective in reducing longer voltage violations while leading to many shorter voltage violations. This results in a large number of cumulative violations. Overall, the stage-2 policy exhibits optimal performance in terms of reducing the total voltage violation duration. Nevertheless, from time to time all PVs have inevitably reached their maximum regulating capability, as shown in Fig. \ref{fig7}. 
\vspace{-0.10in}
\begin{table}[h]
\scriptsize
\caption{Statistics of the Voltage Event }
\vspace{-0.15in}
\begin{center}
\begin{tabular}{ccccc}
\hline
\hline
\textbf{Statistics}&\textbf{Base}&\textbf{Conv.}&\textbf{Stage 1}&\textbf{Stage 2}\\
\hline
Count& 4031	&30219	& 4831 &\textbf{3314}\\
Mean& 6.74	&1.16	&2.64 &\textbf{2.08}	\\
Std&15.83	&1.075	& 4.16&	\textbf{2.86}\\
25 percentile&1	&1 	& 1 &	\textbf{1}\\
50 percentile& 2 &  1 & 1 & \textbf{1}\\
75 percentile&4	&1	& 2 &\textbf{2}	\\
MaxDuration&95	&43	& 47&\textbf{44}\\
Nodes of MaxDuration &2 &5& 1&\textbf{2}\\
Integration Sum&27176	&34940	&12759 &	\textbf{6914}\\
\hline
\hline
\end{tabular}
\label{tab5}
\end{center}
\vspace{-0.15in}
\end{table}
\subsection{Impact of Action Cost}
Because action-taking incurs a cost (specified by  (\ref{eqn5})), we next conduct a sensitivity analysis in $C_w$.  To observe the impact of action cost (i.e., the value of $ w_{\mathrm{cost}}$) on VVC control performance, we present the performance comparison using different $w_{\mathrm{cost}}$ values in three spring days. As shown in Table {\ref{tab6}}, there is a noticeable decrease of $Q_{total}$ when $ w_{\mathrm{cost}} = 0.005$, but the degradation in voltage score seems less evident. However, if $ w_{\mathrm{cost}}$ increases to 0.01, the voltage score declines significantly due to the lack of action from the agents.
\vspace{-0.15in}
\begin{table}[h]
\scriptsize
\caption{Impact of Action Cost on VVC Performance}
\vspace{-0.15in}
\begin{center}
\begin{tabular}{cccccccc}
\hline
\hline
$C_w$& $0.001$&$0.002$&$0.003$&$\textbf{0.005}$&\textcolor{red}{$\textbf{0.01}$}\\
\hline
Voltage Score&0.999935& 0.999935& 0.999777& \textbf{0.999789}&\textcolor{red}{\textbf{0.997897}}\\
$Q_{total}(kVar)$&225.106&224.310&216.976 & \textbf{206.443} &\textcolor{red}{\textbf{72.033}} \\
\hline
\hline
\end{tabular}
\label{tab6}
\end{center}
\vspace{-0.15in}
\end{table}

\section{Conclusion}
In this paper, we develop a two-stage progressive training strategy for improving the training speed and convergence when training multiple RL-based VVC agents in high PV-penetration distribution systems.  
%that could successfully lead to cooperative policy of PV agents. 
Simulation results substantiate that stage-1 training can make agents effectively learn when their actions are effective, while stage-2 training can further strengthen the agents's understanding on how to coordinate with others to achieve satisfactory VVC performance. Most importantly, the policy obtained by the stage-1 can also serve as a backup strategy in case communication may be disconnected. Our follow-up journal paper will present the algorithm in detail with extensive testing results on actual feeder models. 
%o reduce the observation needs and address the difficulties in coordination with heterogeneous control devices in case the regulating capability is not enough.
%provide  the proposed strategy could trained agents that achieve the best voltage performance in terms of voltage score and error reduction with least reactive power cost in worst season. The first stage strategy can make for the communication loss with limited local observations.

\bibliographystyle{IEEEtran}
\bibliography{references2}

\end{document}